\documentclass[%
 reprint,
 amsmath,amssymb,
 aps,
]{revtex4-2}

\usepackage[utf8]{inputenc}
\usepackage{tabularx}
\usepackage{graphicx}
\usepackage{dcolumn}
\usepackage{bm}
\usepackage{amsmath}
\usepackage{amssymb}
\usepackage{subeqnarray}
\usepackage{cases}
\usepackage{enumitem}
\usepackage{xcolor}

\usepackage{esint}
\usepackage{subcaption}

\newcommand{\rmd}{{\rm d}}

\begin{document}

\title{Moving impedance profiles make one--way, spectrum--reshaping mirrors}

\author{Mingjie Li}
\email{ml813@exeter.ac.uk}

\author{S. A. R. Horsley}
\email{S.Horsley@exeter.ac.uk}
\affiliation{Department of Physics and Astronomy, University of Exeter, Stocker Road, Exeter, UK, EX4 4QL}

\vspace{10pt}

\begin{abstract}
We find a new set of exact solutions to Maxwell's equations in space--time varying materials, where the refractive index is constant, while the impedance exhibits effective motion, i.e. it is a function of $x-vt$.  We find that waves co--propagating with the modulation are not reflected within the material, while counter--propagating waves are continually reflected by the changing impedance.  For a finite section of such a material we find analogues of transmission resonances, where specially shaped `eigenpulses' enter without reflection.  We also find that there is a strong asymmetry in reflection from the medium when the impedance modulation is small but rapid, the material reflecting strongly from one side, and negligibly from the other.  Unlike stationary media, the spectrum of the reflected wave can be significantly different from the incident one.
\end{abstract}

\maketitle
%
\vspace{2pc}

Spatially varying electromagnetic and acoustic materials have been a subject of research, even preceding the discovery of waves.  Yet---despite some important exceptions \cite{morgenthaler1958,felsen1970wave,koutserimpas2018electromagnetic,fante1973propagation,holberg1966parametric,budko2009electromagnetic}---comparatively little work has considered temporally, or even space--time varying materials.  Spurred on by experimental developments in this area \cite{zhou2020,tirole2023}, this has begun to change and there is an increasing body of work exploring the potential of these materials \cite{galiffi2022}.  Many interesting properties have already been identified, such as non-reciprocal propagation \cite{nonReciprocalPhotonics}, temporal aiming \cite{pacheco2020}, temporal anti--reflection coatings \cite{pacheco2020b}, and parametric amplification and invisibility based on PT--symmetry \cite{parametricAmp}.

Here we investigate the electromagnetic permittivity and permeability in terms of the two moving coordinates $u=x-vt$ and $w=x+vt$, instead of the space and time variables, $x$ and $t$.  We take the velocities of the moving coordinates, $v$ as constant.  These materials are a subset of space--time modulated materials that exhibit synthetic motion, and have been widely researched \cite{Uniform-velocity,HomoTheoST,SyntheticAxionResponse,Oliner1961wave,Cassedy1963dispersion,Cassedy1967dispersion,pendry2021:gainEnergy,Galiffi2019amplification,FresnelDrag,ArchimedesScrewForLight}. It is already known that they exhibit some novel properties, such as pulse compression and amplification \cite{pendry2021:gainEnergy,Galiffi2019amplification}, synthetic Fresnel drag \cite{FresnelDrag}, an analogue ``Archimedes' screw'' for light \cite{ArchimedesScrewForLight}, and the emission of photons from the vacuum state \cite{horsley2023quantum}.  In much of this work there is no general analytic solution to the wave equation, except in the special case of impedance matched media.

\begin{figure}[h!]
    \includegraphics[width=0.45\textwidth]{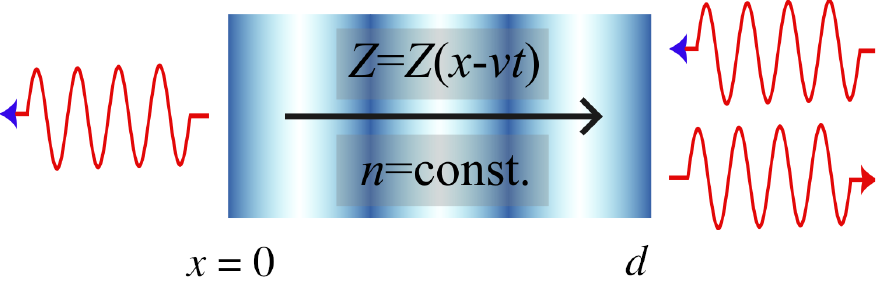}
    \caption{\textbf{Schematic of problem.} A wave propagates through a space-time varying medium extended from $x=0$ to $x=d$, with uniform wave speed and an impedance profile that moves with a velocity equal to the wave speed, $v=c/n$, i.e. $Z=Z(x-vt)$.}
    \label{fig:set-up}
\end{figure}

In this paper, we shall derive a new general solution to the wave in a $u$--variant material whose refractive index is constant but has an impedance that is an arbitrary function of $u=x-vt$, including the effects of reflection at a boundary with another medium e.g. vacuum, as illustrated in figure \ref{fig:set-up}.  We note that finding the solution to the wave equation in a purely $u$--variant material is the same as finding it in a purely $w$--variant one, since these coordinates can be interchanged by flipping the sign of the velocity, $v$.    

%
%

Consider an electromagnetic material that has a constant refractive index, namely, $n=\text{const.}$, but an impedance profile which is moving at velocity $v$, i.e. $Z=Z(x-v t)$.  This case is the counterpart to the uniform impedance, varying index media considered in \cite{pendry2021:gainEnergy,horsley2023quantum}.  Due to the uniform refractive index, the speed of light is a constant and here we explore the case where constant $v$ is equal to the wave speed, $v=c/n$.  For simplicity, we consider the $1$+$1$ dimensional case, assuming that the wave propagates along the $x$ axis and has a uniform amplitude in the $y$--$z$ plane.

As we shall show, such materials have the interesting property that, for waves propagating in the same direction as the moving impedance profile, the impedance at each point in the wave is constant, leading to zero reflection during propagation.  Meanwhile, those waves propagating in opposition to the motion of the profile will be subject to a constantly changing impedance, leading to potentially large reflections.  One may expect this kind of material will work as a one--sided mirror, reflecting the illuminating wave from one direction but barely at all in the opposite direction.  We now show that Maxwell's equations can be solved analytically in this system, for any choice of impedance profile $Z(x-vt)$.

With the above assumptions, Maxwell's equations reduce to
\begin{subequations}\label{eq:MaxwellEqs2D}
\begin{equation}
\partial_x E_y=-\frac1v\partial_t(ZH_z), \label{eq:MaxwellEqs2D_a}
\end{equation}
\begin{equation}
-\partial_x H_z=\frac1v\partial_t\frac{E_y}Z, \label{eq:MaxwellEqs2D_b}      
\end{equation}
\end{subequations}
where $Z=\sqrt{\mu_0\mu/\epsilon_0\epsilon}$ is the impedance of the medium, and we have assumed a polarization such that $E_z=H_y=0$. To solve this pair of equations, we introduce the left and right moving coordinates described in the introduction, $u=x-vt$ and $w=x+vt$.  With respect to this pair of new coordinates, the previous pair of Maxwell's equations (\ref{eq:MaxwellEqs2D}) take the form,
\begin{subequations}\label{eq:MaxwellEqs2D_uw}
\begin{equation}
 \partial_u(E_y-ZH_z)+\partial_w(E_y+ZH_z)=0, \label{eq:MaxwellEqs2D_uw_a}   
\end{equation}
\begin{equation}
-\partial_u\left(\frac1Z(E_y-Z H_z)\right)+\partial_w\left(\frac1Z(E_y+Z H_z)\right)=0. \label{eq:MaxwellEqs2D_uw_b}     
\end{equation}       
\end{subequations}
We have not yet imposed the functional form of the impedance.  Assuming it is a function of $u$ alone, $Z$ can be taken outside all of the $w$ deivatives in Eq. (\ref{eq:MaxwellEqs2D_uw_a}--\ref{eq:MaxwellEqs2D_uw_b}).  This is especially important in Eq. (\ref{eq:MaxwellEqs2D_uw_b}), which can now be written as $\partial_w(E_y+Z H_z)=Z\partial_u(Z^{-1}(E_y-ZH_z))$.  We can then apply Eq. (\ref{eq:MaxwellEqs2D_uw_a}) to completely eliminate the $w$ derivative, reducing the problem to a differential equation in a single variable
\begin{equation}
\partial_u \ln\left(\frac{(E_y-ZH_z)}{\sqrt Z}\right)=0.
\end{equation}
This tells us that the ratio of the part of the field incident from the right, $E_{y}-Z H_{z}$, and the square root of the impedance does not depend on the $u$ coordinate. The general solution to this equation is such that the argument of the logarithm equals an arbitrary function of the second moving coordinate, $w$, i.e.
\begin{equation} \label{eq:solution_E-ZH}
    E_y(u,w)-ZH_z(u,w)=\sqrt{Z}g(w),
\end{equation}
To find left--incident part of the electromagnetic field, $E_y+Z H_z$ we can now substitute (\ref{eq:solution_E-ZH}) into Eq. (\ref{eq:MaxwellEqs2D_uw_a}) and integrate with respect to $w$, finding
\begin{equation}\label{eq:solution_E+ZH}
    E_y(u,w)+ZH_z(u,w) =-G(w)\partial_u\sqrt{Z}+ h(u)
\end{equation}
where $G(w)=\int^{w} g(w') \rmd w'$ and $h(u)$ is an arbitrary function of $u$.  We can observe from Eqs. (\ref{eq:solution_E-ZH}) and (\ref{eq:solution_E+ZH}) that when $g=0$, the wave is purely right--going within the material, being a function of the variable $u$ alone, obeying the expected proportionality between electric and magnetic fields $E_y= Z H_z$.  Right--going waves are therefore not reflected as they propagate through this medium, even though there is a spatially and temporally varying impedance.  As indicated above, this is because each point in the moving wave is at a fixed value of the impedance: the ratio between the electric and magnetic fields vary throughout the wave, but this variation is carried with the wave, leading to zero reflection.

Meanwhile, setting $h(u)=0$ in Eqns. (\ref{eq:solution_E-ZH}) and (\ref{eq:solution_E+ZH}), the electromagnetic wave becomes a function of both variables $u$ and $w$, indicating it is now a combination of left and right going parts.  In regions where the impedance $Z$ is constant, $E_y=-Z H_z$, and the wave reduces to being purely left--going.  We therefore interpret the arbitrary function $g$ as specifying the combination of left and right--going fields within the medium, due to a wave incident from $+\infty$.

Adding and subtracting Eqs. (\ref{eq:solution_E-ZH}--\ref{eq:solution_E+ZH}), we can find expressions for the electric and magnetic fields that are the general solution to (\ref{eq:MaxwellEqs2D}),
    \begin{equation}
        E_y =\frac12\left[-\left(\sqrt{Z(u)}\right)'G(w)+\sqrt{Z(u)}G'(w) +h(u)\right],\label{eq:GeneralSolution_E}
    \end{equation}
    and
    \begin{equation}
        H_z =\frac{1}{2Z(u)}\left[-\left(\sqrt{Z(u)}\right)'G(w)-\sqrt{Z(u)}G'(w)+h(u)\right].\label{eq:GeneralSolution_H}
    \end{equation}    
As stated above, we interpret the arbitrary function $h$ as specifying the amplitude of waves within the medium, incident from $-\infty$.  As these waves move in lock step with the impedance profile, no reflected wave is generated.  We similarly interpret the arbitrary function $g$ as specifying the amplitude of waves incident from $+\infty$.  This part of the field is continuously incident onto a space and time varying impedance, generating significant reflection.
%
%
\begin{figure}[h!]
    \begin{subfigure}[h]{0.45\textwidth}
        \includegraphics[width=\textwidth]{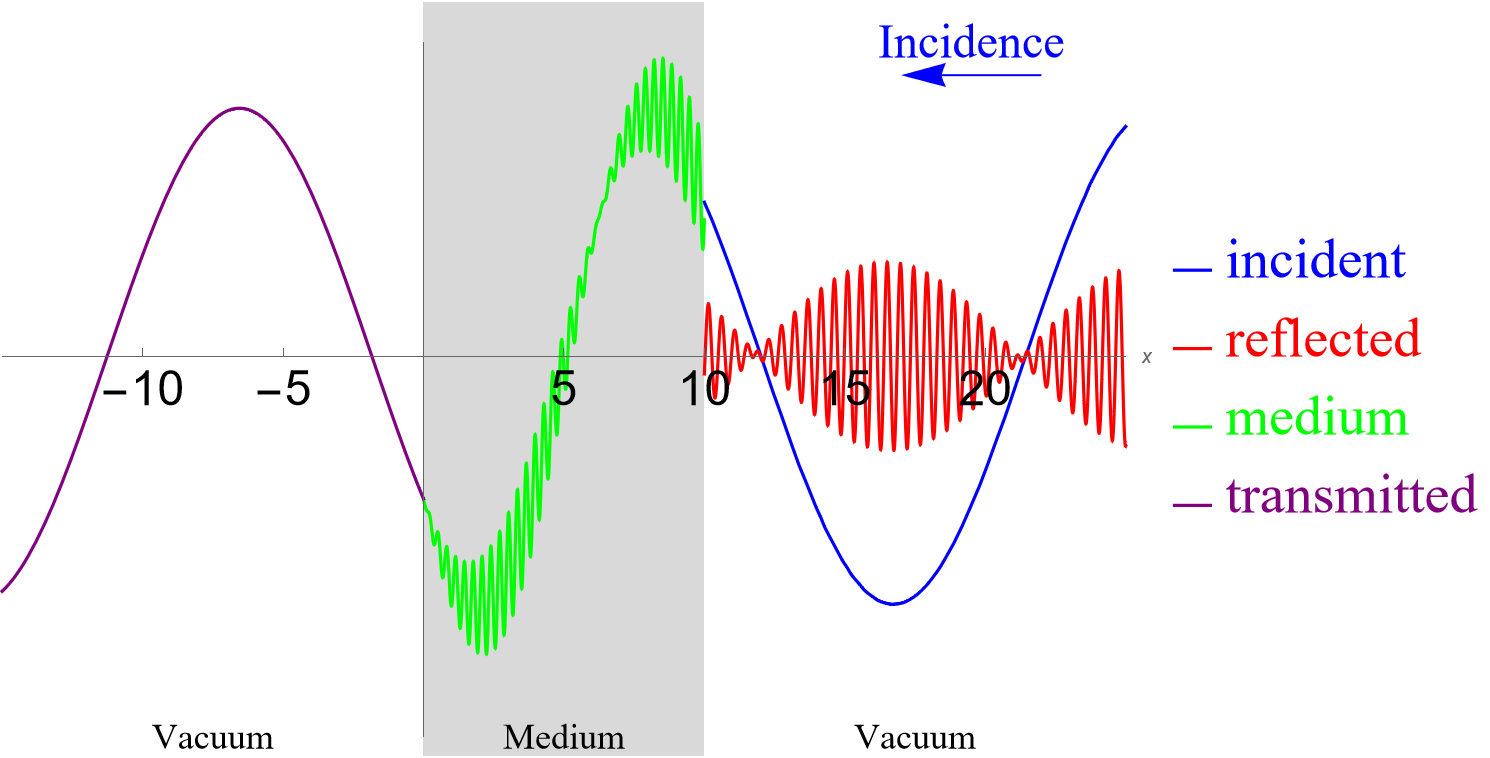}
        \caption{Incidence from the right, where reflection is not negligible.  The arbitrary function $G$ defined in Eqs. (\ref{eq:GeneralSolution_E}--\ref{eq:GeneralSolution_H}) is here equal to $G(w)=6\sin(0.5w)$. }
        \label{fig:right_incident}
    \end{subfigure}
    \begin{subfigure}[h]{0.45\textwidth}
        \includegraphics[width=\textwidth]{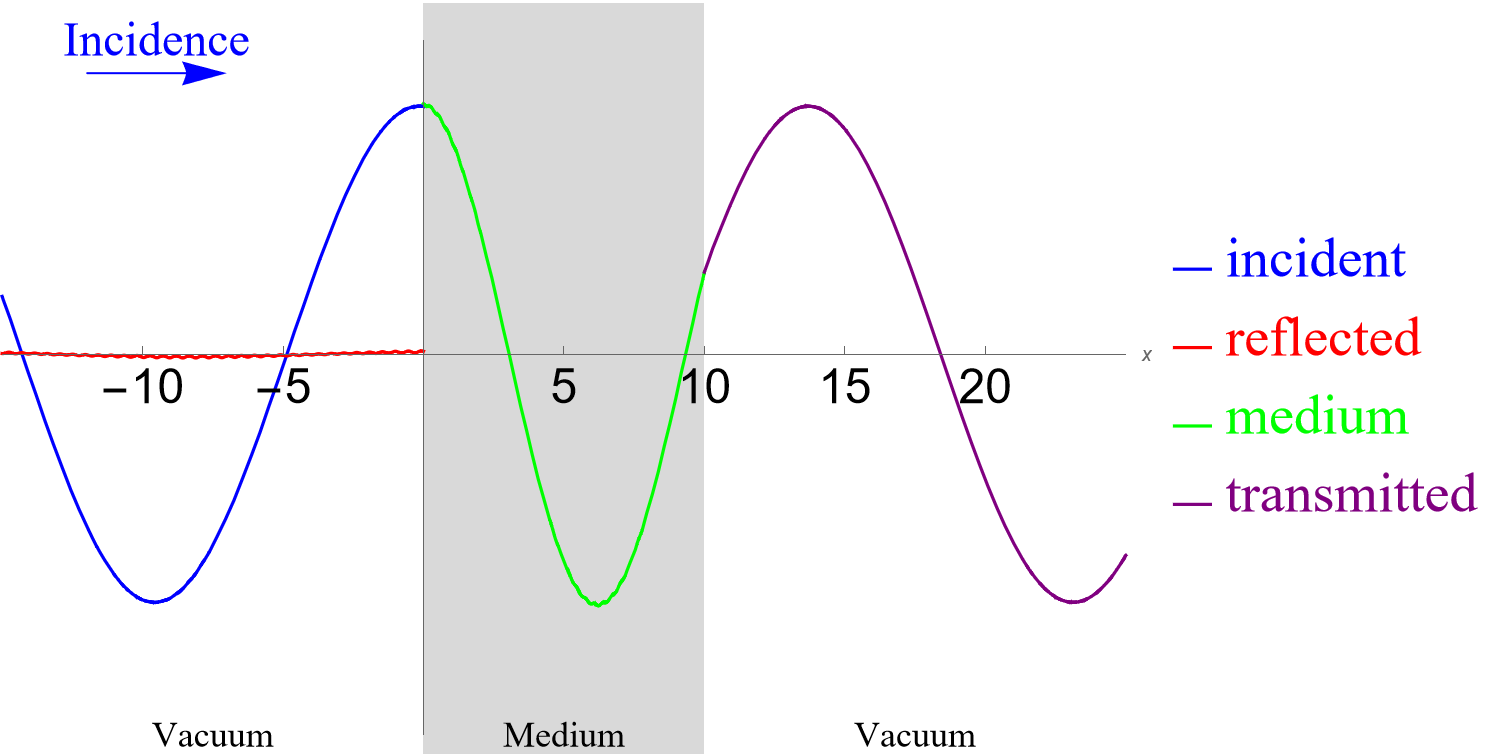}
        \caption{Incidence from the left, where reflection is negligible. Here $G(w)$ is the numerical solution to $G'(w)=-\frac{Z(2x_0-w)/\eta_0-1}{Z(2x_0-w)/\eta_0+1}\times3\cos(0.5w)$, so that the transmitted wave function is the same as that above. }
        \label{fig:left_incident}
    \end{subfigure}    
    \caption{\textbf{Asymmetric propagation through a weakly modulated impedance profile}. Figures (a) and (b) show the difference between waves incident from the left and right onto a moving impedance profile $Z(x-vt)$ where $v=c/n$.  Here the impedance is weakly modulated and takes the form $Z(u)/\eta_0=1.011+0.01\cos(20 u)$, while the refractive index takes the fixed value 1.5 (indicated as the shaded region).  Outside the shaded region we assume vacuum.}
    \label{fig:example_one_side_mirror}
\end{figure}

Using the above expressions for the electric and magnetic fields (\ref{eq:GeneralSolution_E}--\ref{eq:GeneralSolution_H}), the electromagnetic energy distribution is given by the usual expression $\mathcal{E}=\frac12\varepsilon E_y^2+\frac12\mu H_z^2$, which equals
\begin{multline}\label{eq:EnergyDensity}
     \mathcal{E}=\frac n{4cZ}\bigg[ \left(\sqrt Z G'\right)^2+\left(\left(\sqrt Z\right)'G\right)^2  + h^2\\
     - 2\left(\left(\sqrt Z\right)' G\right)h \bigg].
\end{multline}
Using the same substitution, the corresponding Poynting vector is equal to
\begin{align}\label{eq:Poynting}
S&=E_y H_z\nonumber\\[10pt]
&= \frac1{4Z} \left[ -(\sqrt Z G')^2+((\sqrt Z)'G)^2 + h^2- 2(\sqrt Z)' G h \right].
\end{align}
As the Poynting vector (\ref{eq:Poynting}) is the current of energy flow, the ratio of the Poynting vector to the energy density equals the velocity of energy flow, $S/\mathcal{E}=\mathcal{U}$, which is
\begin{equation}
    \mathcal{U}=\frac{c}{n}\left[\frac{((\sqrt Z)'G)^2 -(\sqrt Z G')^2 + h^2- 2(\sqrt Z)' G h}{((\sqrt Z)'G)^2 + (\sqrt Z G')^2  +h^2-2((\sqrt Z)' G)h }\right]\label{eq:energy-velocity}
\end{equation}
This expression for the energy velocity confirms the above interpretation of the solutions (\ref{eq:GeneralSolution_E}) and (\ref{eq:GeneralSolution_H}), where right--going waves are unreflected within the medium, while left--going ones are not.  In general the energy flow is can be positive or negative, and changes in space and time.  However, setting $G=0$ in (\ref{eq:energy-velocity}), the energy velocity reduces to the constant positive value $\mathcal{U}= c/n$, i.e. the flow of energy in a right--going wave propagates at a constant velocity, despite the space--time variation of the impedance.  Meanwhile, setting $h=0$, the energy velocity reduces to $\mathcal{U}=(c/n)[(\sqrt{Z}'G)^2-(\sqrt{Z}G')^2]/[(\sqrt{Z}'G)^2+(\sqrt{Z}G')^2]$, which equals $-c/n$ when the impedance is uniform.  We can interpret the two contributions, $\sqrt{Z}'G$ and $\sqrt{Z}G'$ as the reflected and transmitted parts of the energy flow, respectively.  In general these combine such that the energy velocity varies in space and time and can take either sign, indicating that left--going waves are constantly reflected within such a moving impedance profile.  In cases where $|\sqrt{Z}'G|>|\sqrt{Z}G'|$ this reflection is strong enough to reverse the energy flow, making $\mathcal{U}$ positive despite the wave incoming from $+\infty$. 

The above analysis shows that wave propagation in a medium with constant wave speed, but a moving impedance profile is highly asymmetric, leaving one direction of propagation unaffected while significantly reflecting the other, potentially with a diode--like function.  Yet, the solutions we have found are only valid in a medium with an everywhere continuously varying impedance profile and a constant wave velocity.  In practice we must impose the boundary conditions that the material becomes vacuum at large $|x|$.  One way to do this is to consider a function $Z$ that tends to a constant at large distances, although this restricts us to $n=1$ everywhere.  It is thus more realistic to consider an abrupt interface with vacuum, requiring us to find the superposition of left and right going--waves that make the electric and magnetic fields continuous at the interfaces.

A straightforward way to approach the problem of reflection from a finite section of such a time varying material is to calculate the combinations of fields $E_{\pm}=\frac{1}{2}\left(E_{y}\pm\eta_0 H_z\right)$ within the medium.  In vacuum these combinations respectively represent the right and left--going parts of the field, and given the continuity of $E_y$ and $H_z$ these represent the right and left--going fields at the interfaces of the medium with vacuum.  Substituting our general solutions (\ref{eq:GeneralSolution_E}--\ref{eq:GeneralSolution_H}) we find,
\begin{multline}
    E_{\pm}=\frac{1}{4}\bigg[\left(1\mp\frac{\eta_0}{Z(u)}\right)\sqrt{Z(u)}G'(w)\\[10pt]
    +\left(1\pm\frac{\eta_0}{Z(u)}\right)\left(h(u)-\left(\sqrt{Z(u)}\right)'G(w)\right)\bigg].\label{eq:Epm}
\end{multline}
For a wave incident from the left hand side of the medium, the left--going part of the field must be zero on the right hand interface at $x=d$, i.e. $E_{-}(x=d)=0$.  This implies the following relationship between the arbitrary functions $h$ and $G$
\begin{multline}
    \text{Left incidence:}\qquad h(u)=\left(\sqrt{Z(u)}\right)'G(2d-u)\\[10pt]
    -\left(\frac{1+\frac{\eta_0}{Z(u)}}{1-\frac{\eta_0}{Z(u)}}\right)\sqrt{Z(u)}\,G'(2d-u)\label{eq:h-left-incidence}
\end{multline}
Similarly for incidence from the right hand side, the right--going part of the field must reduce to zero on the left hand boundary at $x=0$, i.e. $E_{+}(x=0)=0$.  This requirement fixes the function $h(u)$ in the same way as (\ref{eq:h-left-incidence}),
\begin{multline}
    \text{Right incidence:}\qquad h(u)=\left(\sqrt{Z(u)}\right)'G(-u)\\[10pt]
    -\left(\frac{1-\frac{\eta_0}{Z(u)}}{1+\frac{\eta_0}{Z(u)}}\right)\sqrt{Z(u)}\,G'(-u).\label{eq:h-right-incidence}
\end{multline}
These relations, (\ref{eq:h-left-incidence}) and (\ref{eq:h-right-incidence}) complete the solution to the problem, fixing the incident, reflected, and transmitted fields in terms of the arbitrary function $G$, for incidence from both the left and right hand side of the medium.   Yet as it stands, the physical meaning of our solution is not straightforward to understand.  This is because we must specify the form of the field \emph{inside} the medium and only then can we determine the form of the incident and scattered waves.  Specifying the solution in terms of the incident wave would be more useful.

One approach, similar to that explained in \cite{horsley2023}, is to re--write Eqs. (\ref{eq:Epm}--\ref{eq:h-right-incidence}) in terms of operators acting on the unknown function $G$.  From this the reflection and transmission operators can be derived, determining the transformation of the incident pulse enacted by the medium.  However we do not derive those operators here, for simplicity here we instead simply specify our solutions in terms of the form of the field on the \emph{transmission} side of the medium.  Substituting Eqs. (\ref{eq:h-left-incidence}--\ref{eq:h-right-incidence}) into (\ref{eq:Epm})
and evaluating the appropriate fields on the transmission side of the medium we find the function $G$ is determined by the transmitted field, $E_{\rm t}$, via the simple relations,
\begin{multline}
\text{Left incidence:}\qquad E_{\rm t}(t)=\frac{\sqrt{Z(d-vt)}}{1-\frac{Z(d-vt)}{\eta_0}}G'(d+vt)\label{eq:Gp_left}\\
\end{multline}
and,
\begin{multline}
\text{Right incidence:}\qquad E_{\rm t}(t)=\frac{\sqrt{Z(-vt)}}{1+\frac{Z(-vt)}{\eta_0}}G'(vt),\label{eq:Gp_right}\\
\end{multline}
where $E_{\rm t}$ is the transmitted field evaluated on the exit face of the medium ($x=d$, for left incidence, $x=0$ for right).  The fields plotted in Fig.~\ref{fig:example_one_side_mirror} were computed in this way: through specifying the transmitted fields $E_{\rm t}$, finding the function $G$ via Eqs. (\ref{eq:Gp_left}--\ref{eq:Gp_right}), then evaluating the incident and reflected fields on the entrance face of the medium, from Eqs. (\ref{eq:Epm}--\ref{eq:h-right-incidence}).  The explicit expressions for the incident and reflected fields are rather lengthy and are given in the appendix below.

\begin{figure}[h!]
    \centering
    \includegraphics[width=0.45\textwidth]{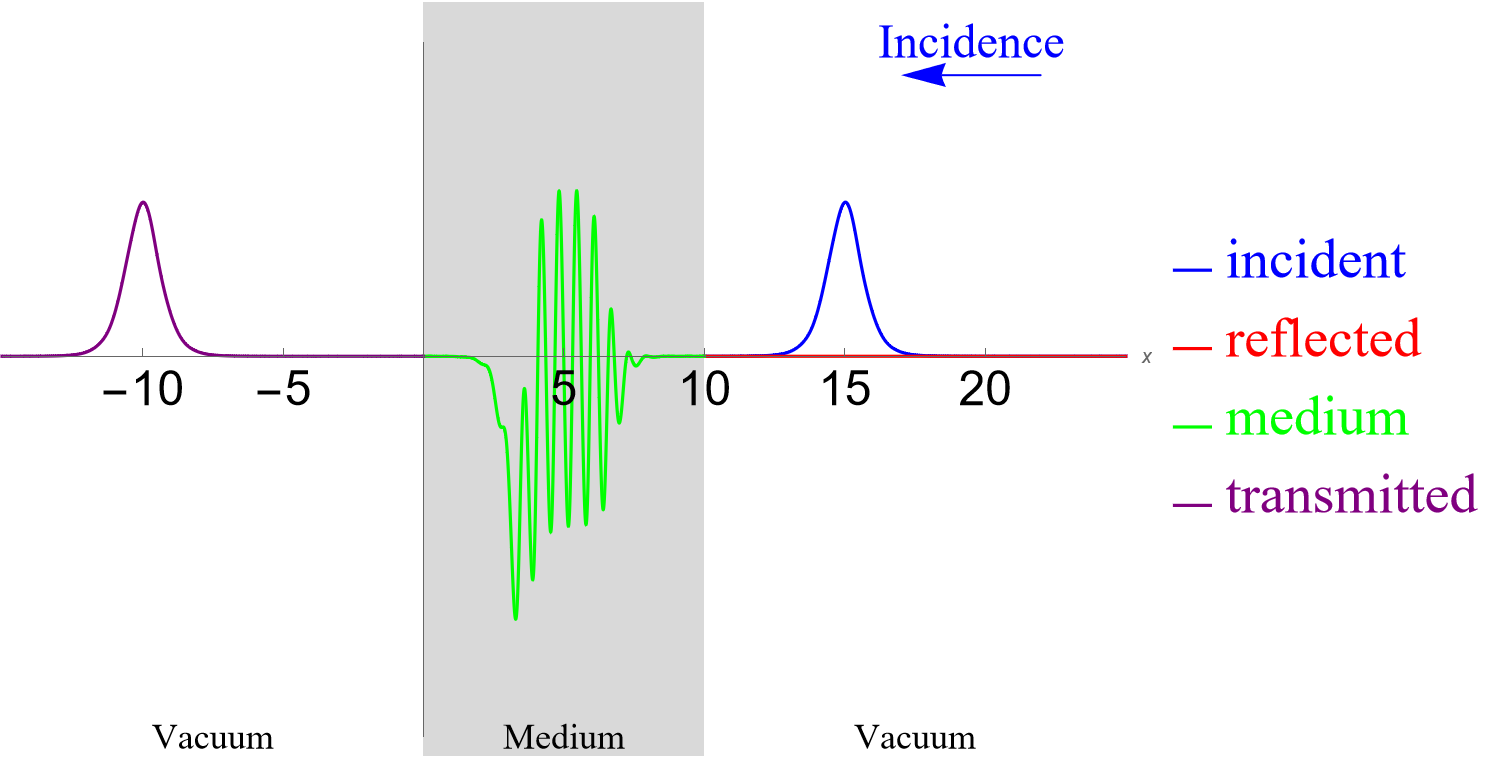}
    \caption{\textbf{Reflectionless Eigenpulses}. Any choice of $G$ that is periodic with period $2d$ corresponds to an incident wave--form leading to zero reflection.  The above example shows this effect for $G(w)=\tanh(6\sin(\pi w/d))$ and an impedance $Z(u)=1.5+0.25\cos(10 u)$, corresponding to a sequence of hyperbolic secant--like incident pulses.}
    \label{fig:reshaping}
\end{figure}

There are some interesting special cases where the above exact solution can be simplified.  First is the case where the function $G$ is periodic, with period $2d$.  As is evident in formulae (\ref{eq:Er_L}) and (\ref{eq:Er_R}) given in the appendix, in this case the reflection vanishes for incidence from either side of the medium.  The transmitted fields remain equal to the above expressions (\ref{eq:Gp_left}--\ref{eq:Gp_right}), whereas the incident fields (\ref{eq:Ei_L}) and (\ref{eq:Ei_R}) reduce to $E_{\rm i}=\sqrt{Z(-vt)}(1-Z(-vt)/\eta_0)^{-1}G'(vt)$ and $E_{\rm i}=\sqrt{Z(d-vt)}(1+Z(d-vt)/\eta_0)^{-1}G'(d+vt)$ for left and right incidence respectively.  For the case of a constant impedance this vanishing reflection is simply a transmission resonance, where the reflection from the front interface destructively interferes with that from the back, making the net reflection vanish.  For a non--uniform, space--time varying impedance the vanishing reflection is also a kind of transmission resonance.  However, instead of the usual condition of zero reflection for a fixed frequency, here the vanishing reflection occurs for the special incident wave--forms given above (see Fig. \ref{fig:reshaping}).  These are simple analytic examples of the zero reflection eigenpulses discussed in Ref.~\cite{horsley2023}.  Interestingly---unlike the case of a uniform impedance---the transmitted and reflected fields are not related by a simple time translation, showing that the medium also re--shapes these fields in transmission.
%
%

Another interesting limit is one where the impedance is only very weakly modulated.  This is of most practical interest as small, rapid material modulations are easier to realise than large amplitude modulations.  We thus take the impedance to have the form, $Z/\eta_0=1+\delta\eta$, where $|\delta\eta|\ll1$.  In this case the incident and reflected fields (\ref{eq:Er_L}--\ref{eq:Er_R}) reduce to a simple form.  For incidence from the left we have,
\begin{align}
    E_{\rm i}&\sim-\frac{\sqrt{\eta_0}}{\delta\eta}\left(1+\frac{\delta\eta}{2}\right) G'(2d+vt)\nonumber\\
    E_{\rm r}&\sim\frac{\sqrt{\eta_0}}{2}\left[G'(vt)-G'(2d+vt)\right]\nonumber\\
    E_{\rm t}&\sim-\frac{\sqrt{\eta_0}}{\delta\eta}\left(1+\frac{\delta\eta}{2}\right)G'(d+vt).\label{eq:Eirt_L}
\end{align}
and similarly for incidence from the right,
\begin{align}
    E_{\rm i}&\sim\frac{\sqrt{\eta_0}}{2}G'(d+vt)
    \nonumber\\
    E_{\rm r}&\sim\frac{\sqrt{\eta_0}}{4}\bigg[\delta\eta\left(G'(d+vt)-G'(-d+vt)\right)\nonumber\\
    &\qquad\qquad+\delta\eta'\left(G(-d+vt)-G(d+vt)\right)\bigg]\nonumber\\
    E_{\rm t}&\sim\frac{\sqrt{\eta_0}}{2}G'(vt)\label{eq:Eirt_R}
\end{align}
where we have taken all quantities to leading plus next order in $\delta\eta$, and $\delta\eta$ is evaluated at the same point in space as the field component.  As expected, for both directions of incidence, the incident and transmitted fields are of the same order in the impedance contrast $\delta\eta$, whereas the reflected field is of the next order in smallness.  This guarantees that the reflection from the medium vanishes as $\delta\eta\to0$, whereupon the incident and transmitted fields become identical, but delayed by the propagation time $d/v$.  In addition, as the thickness of the medium $d$ is reduced to zero, the reflected field also vanishes and the incident and transmitted fields become identical.  Although the sets of fields (\ref{eq:Eirt_L}) and (\ref{eq:Eirt_R}) are very similar, there is one crucial difference: for left incidence the reflected field is of order $\delta\eta$ times the incident field, whereas for right incidence there is a contribution of order $\delta\eta'$ times the integral of the incident field.  This means that for e.g. an ultra--low contrast but rapidly varying impedance contrast $\delta\eta$, the reflection from the right of the medium will be much larger than that from the left. Moreover, the reflected field inherits the rapid variation of the impedance and the incident spectrum is thus converted to much a much higher frequency reflected one.  The medium is thus effectively transparent for incidence from the left, whereas it acts as a frequency up--converting mirror for right--hand incidence.  This effect is illustrated in Fig. \ref{fig:example_one_side_mirror}. 

%
%
To summarise, we have found that Maxwell's equations are exactly soluble in space--time varying media where the refractive index is constant, while the impedance profile moves at the wave speed $v=c/n$.  This adds to existing analytic results that establish exact solutions in synthetically moving, impedance matched media~\cite{pendry2021:gainEnergy,horsley2023quantum}.  Our results are complementary to this existing work, which assumes zero reflection, and where propagation is dominated by changes to the wave speed.  In this earlier work it was found that when the wave speed equals the modulation velocity there is an intense compression of the field, analogous to a travelling wave amplifier.  By contrast, here we've shown that when the main effect of the moving medium is to reflect the wave rather than change the wave speed, there is no such pulse compression.  This distinction is straightforward to understand through examining the ray diagrams shown in e.g. Ref. \cite{horsley2023quantum}.

In this family of space--time varying materials, wave propagation is quite different.  In general the reflection and transmission is asymmetric, being different for the two directions of incidence.  The reflected and transmitted fields are also reshaped, inheriting the time modulation of the impedance.  Interestingly, when the modulation contrast is small but rapid, we have shown in Eqs. (\ref{eq:Eirt_L}--\ref{eq:Eirt_R}) that the reflection from one side of the medium becomes negligible, whereas from the other it is larger by a factor proportional to the rate of change of the impedance.  Moreover the reflected wave inherits the time modulation of the impedance and is thus shifted to a much higher frequency compared to the incident wave. 

Another interesting effect we have observed in our solution is the existence of eigenpulses (see Ref. \cite{horsley2023}) that are not reflected from the medium, irrespective of the impedance contrast.  As discussed above, we have shown that these are the analogues of transmission resonances that occur in uniform media, where the wave inside the medium has a time variation such that it has the same form on the entrance and exit facets of the medium.

In addition to illuminating the role of reflection in synthetically moving media, it should be possible to observe these effects in modulated radio frequency transmission lines as was recently done in e.g. \cite{moussa2023}.  Similar effects should also be possible to realise in acoustics using e.g. electronic feedback~\cite{wen2022} or the various implementations of time modulation reviewed in \cite{galiffi2022}.

\section*{Acknowledgements}
Mingjie thanks the China Scholarship Council for financial support.  SARH thanks J. B. Pendry for illuminating conversations and the Royal Society and TATA for financial support (RPG-2016-186), as well as support through EPSRC program grant ``Next generation metamaterials: exploiting four dimensions'' EP/Y015673/1. 

\section*{Appendix}
Here we give the full expressions for the incident and reflected fields, in terms of the arbitrary function $G$ determined by Eqs. (\ref{eq:Gp_left}--\ref{eq:Gp_right}), for the two directions of incidence.

\begin{widetext}
For incidence from the left, substituting the form of $h(u)$ from Eq. (\ref{eq:h-left-incidence}) into the left and right--going parts of the field (\ref{eq:Epm}) on the entrance interface at $x=0$ gives and incident field of the form,
\begin{multline}
    E_{\rm i}=\frac{1}{4}\bigg\{\frac{\sqrt{Z(-vt)}}{1-\frac{\eta_0}{Z(-vt)}}\left[\left(1-\frac{\eta_0}{Z(-vt)}\right)^2 G'(vt)-\left(1+\frac{\eta_0}{Z(-vt)}\right)^2 G'(2d+vt)\right]\\
    +\left(\sqrt{Z(-vt)}\right)'\left(1+\frac{\eta_0}{Z(-vt)}\right)\left(G(2d+vt)-G(vt)\right)\bigg\}\label{eq:Ei_L}
\end{multline}
and a reflected field,
\begin{multline}
    E_{\rm r}=\frac{1}{4}\bigg[\sqrt{Z(-vt)}\left(1+\frac{\eta_0}{Z(-vt)}\right)\left(G'(vt)-G'(2d+vt)\right)+\left(\sqrt{Z(-vt)}\right)'\left(1-\frac{\eta_0}{Z(-vt)}\right)\left(G(2d+vt)-G(vt)\right)\bigg].\label{eq:Er_L}\\
\end{multline}
Note the appearance of both the function $G(vt)$ and its shifted counterpart $G(2d+vt)$, which represent the contributions to the field from the internal waves at the two separate interfaces.  When the function $G$ is unchanged after the time delay $2d/v$ then the internal field is the same on both interfaces, effectively shrinking the medium to zero width, whereupon the reflection vanishes.

Similarly, for incidence from the right we substitute the form of $h(u)$ from Eq. (\ref{eq:h-right-incidence}) into (\ref{eq:Epm}) on the entrance interface at $x=d$, which gives incident and reflected fields of the form
\begin{multline}
    E_{\rm i}=\frac{1}{4}\bigg\{\frac{\sqrt{Z(d-vt)}}{1+\frac{\eta_0}{Z(d-vt)}}\left[\left(1+\frac{\eta_0}{Z(d-vt)}\right)^2 G'(d+vt)-\left(1-\frac{\eta_0}{Z(d-vt)}\right)^2 G'(-d+vt)\right]\\
    +\left(\sqrt{Z(d-vt)}\right)'\left(1-\frac{\eta_0}{Z(d-vt)}\right)\left(G(-d+vt)-G(d+vt)\right)\bigg\}\label{eq:Ei_R}
\end{multline}
and
\begin{multline}
    E_{\rm r}=\frac{1}{4}\bigg[\sqrt{Z(d-vt)}\left(1-\frac{\eta_0}{Z(d-vt)}\right)\left(G'(d+vt)-G'(-d+vt)\right)\\
    +\left(\sqrt{Z(d-vt)}\right)'\left(1+\frac{\eta_0}{Z(d-vt)}\right)\left(G(-d+vt)-G(d+vt)\right)\bigg].\label{eq:Er_R}
\end{multline}
Note that again the reflected field vanishes when $G(vt)=G(2d+vt)$.
\end{widetext}
\bibliographystyle{unsrt}
\bibliography{Ref.bib}
\end{document}